# AI Diffusion to Low- and Middle Income Countries

## A Blessing or a Curse?

Rafael Andersson Lipcsey[†]

July 16, 2025

**Abstract**

Rapid advances in AI have incited extensive inquiry into its effects on productivity and labor, potentially profound in both positive and negative ways. Often neglected, however, is comprehension of how AI technologies diffuse across and within economies. Developing nations, in particular, face substantial labor market impacts from either swift AI adoption or diminished competitiveness from sluggish diffusion. This paper examines the literature on technology diffusion and proposes a tripartite framework to elucidate AI diffusion pathways: global value chains, research collaboration, and inter-firm knowledge transfers. Employing these metrics, it evaluates AI diffusion in sixteen lower- and middle-income countries (LMICs) relative to four developed nations and assesses their dependency on the USA and China. Findings reveal a notable gap in AI diffusion between developed and developing economies, though this chasm is gradually closing. China emerges as a vital source of future diffusion via value chains, while the USA wields greater influence through research and knowledge transfers. Limitations include the exclusion of certain data sources and regions, and the absence of quantitative analysis on diffusion's relationship with technology intensity. Nonetheless, the research surfaces critical macro-level considerations about AI diffusion. It advocates mechanisms to redistribute AI-induced economic gains and bilateral agreements to complement international accords, thereby addressing the diverse needs and risks of economies entering an AI-dominated era. Future inquiries should explore the nexus between AI diffusion, technology intensity, and productivity; refine diffusion measurement methods; incorporate case studies and targeted policy recommendations; and delve deeper into LMIC-specific labor market outcomes.

[†] Email: andlip.rafael@gmail.com

# 1: Introduction

According to the United Nations Conference on Trade and Development, citizens of developing nations currently constitute 83 percent of the world's population, a figure projected to rise to 86 percent by the year of 2050 (UNCTAD 2022). These countries, characterized by fragile labor markets and political systems generally exhibit lower resilience to large-scale structural disruptions engendered by transformative technologies, of which AI is a significant contender.

In coming decades, developing countries will confront the balancing act of regulating the rate at which AI technologies are disseminated in their economies. A sluggish diffusion of AI technologies, described as a process through which innovations spread through a nation's economy, could precipitate a decline in economic competitiveness, increasing the gap in relation to developed nations, induce economic instability, and, in the long term, political and social unrest. Conversely, an overly rapid diffusion rate may instigate abrupt and extensive structural changes, causing for instance a surge in unemployment, challenging these economies' capacity to adapt in a timely manner.

Yet despite these considerations, developing nations are often neglected in AI impacts research.
This paper endeavors to bridge this gap by advancing theoretical understanding of AI technology diffusion and providing empirical insights into diffusion pathways and rates in low- and middle-income countries (LMICs). This research also aspires to inform and refine more precise and targeted policy interventions in the future.

# 2: Literature Review

A foundational step in comprehending the diffusion of AI technologies is to precisely define technology diffusion itself. In essence, technology diffusion can be described as the process through which innovations permeate a nation's economy (Andergassen et al., 2017) and extend internationally through mechanisms such as trade and knowledge sharing (Ferrier et al., 2016; Afonso and Ana Maria Bandeira, 2013; Petsas, 2003). It is important to draw a distinction between diffusion and productivity gains. The latter, reflected for instance through increases in total factor productivity, are expected results of technology diffusion, rather than measures of diffusion itself.

## 2.1: Diffusion through global value chains

It is not surprising, then, that a key pathway for technology diffusion identified in the literature is through global value chains (Castellani et al., 2022; Garcia, 2021; Ferrier et al., 2016). These not only facilitate the trading of products but also serve as avenues for interconnected economies to pool and share know-how, creating value added in the process (The World Bank, 2024).

More specifically, deeper integration into global value chains inherently provides external incentives for companies to adopt new technologies due to increased competition on the global stage (Criscuolo and Timmis, 2017). This heightened competition drives a greater pursuit of efficiency, often supported by the novel technologies themselves (Yang et al., 2021). It has been theorized that this applies to AI as well, which is expected to enhance logistics prediction software and translation capabilities, alleviating numerous bottlenecks that have historically impeded international trade (Jayathilaka, 2022; Ferencz et al., 2022; Achar, 2019).

Further strengthening the connection between international trade and technology diffusion are findings that countries more deeply embedded in global value chains exhibit higher technology intensities, which essentially means that novel technologies are adopted and utilized at a higher rate in these countries (Ferrier et al., 2016).

Lastly, an interesting duality in technology diffusion is uncovered, where both backwards and forward participation in global value chains (receiving foreign value added through importing, and exporting domestic value added) act as complimentary pathways for facilitating technology diffusion (Rigo, 2020).

In general, deepened global value chain (GVC) integration, has been found to yield several benefits, including increased knowledge spillovers (Criscuolo and Timmis, 2017; Ferrier et al., 2016), enhanced supply chain efficiency (Yang et al., 2021), and improved productivity (Criscuolo and Timmis, 2017; Pahl and Timmer, 2019; Rigo, 2020). Furthermore, although supported by a limited number of AI-specific studies (Ding and Dafoe, 2023; Brynjolfsson et al., 2017), there is a general consensus that general-purpose technologies, such as AI, are linked to economic growth, albeit often with a significant time lag (Strohmaier and Rainer, 2016; Jovanovic and Rousseau, 2005).

Current research on AI diffusion is thus more focused on sectoral predictions, such as that certain sectors, like the manufacturing sector, will be much more quickly and largely affected, as it uses for instance sensors, IoT4 devices and other analytics to function in the most efficient way possible (Andersson Lipcsey, 2023). Additionally, process industries like agriculture, chemicals, metals, and mining, which may not produce many direct inputs for AI, still benefit indirectly. Their outputs move through value chains to manufacturing sectors that supply AI with inputs, thus benefiting more from AI diffusion. These sectors also gain from technologies such as AI-powered demand sensing (Andersson Lipcsey, 2023).

Overall however, a picture emerges where AI innovations are most prevalent in sectors that are also the most tradable and deeply integrated into global value chains. According to Ferencz et al. (2022), sectors with the highest foreign value-added in final demand, a measure of global value chain integration, also have the highest shares of patents, trademarks, and publications in AI innovations.

Based on the above, it can thus be concluded that AI diffusion through GVCs is likely to lead to measurable productivity benefits. However, the observed time lag and the lack of up-to-date GVC integration data make this effect difficult to measure at an economy-wide level at present. Nonetheless, attempting to measure the initial step in this process, the diffusion of AI technologies itself, is considered a worthwhile endeavor, as it may provide indicators of expected growth in the future.

### 2.2: Diffusion through knowledge flows

Diffusion through knowledge flows, although sometimes overlapping with processes observed in international trade, can be distinguished as a more immediate avenue for the global spread of AI technologies. This route, also referred to as a type of disembodied technology diffusion, focuses on the dissemination of knowledge without the need for the transfer of goods or services and has been identified as a significant pathway for the percolation of AI technologies (Tang et al., 2022; Jiang et al., 2021; Zhao Rongying et al., 2019; Shih and Chang, 2009).

Three primary avenues for knowledge flows can be outlined: inter-firm knowledge flows, patents, and academic research. Inter-firm flows typically occur most intensely among firms participating in knowledge clusters (Jiang et al., 2021; Andergassen et al., 2017; Alghamdi, 2019). These clusters are not necessarily defined by geographical proximity. Instead, in an increasingly interconnected world due to advancements in ICT, technological proximity, the similarity in technological principles used by firms, plays a similarly, if not more, important role (Andergassen et al., 2017). This view, however, is debated (Tang, Li, et al., 2022). Nonetheless, increased cooperation within knowledge clusters and networks reduces technological distances between firms, thereby increasing knowledge spillovers (Alghamdi, 2019), akin to the effects observed with deepened GVC integration.

Patents are also seen as significant avenues for knowledge flows (Jiang et al., 2021; Zhao Rongying et al., 2019; Shih and Chang, 2009). Diffusion through patent citations can be bidirectional, similar to GVC participation (Rigo, 2020). Institutions or countries can act as technology providers by having their patents cited by others, technology receivers by primarily citing others' patents, or both (Zhao Rongying et al., 2019). However, patents are often viewed more as a pre-proxy for economic growth rather than direct avenues for diffusion (Daniela Di Cagno et al., 2014). In essence, patents are tangible results of diffusion that occur slightly later in the process and serve as better indicators of productivity.

Research networks constitute a third path for diffusion (Tang et al., 2022; Tang, Li, et al., 2022; Ferrier et al., 2016; Shih and Chang, 2009). Although not always elaborated in great detail, the theory posits that co-publication of research papers in the AI field is a preliminary step in technology diffusion, eventually leading to more concrete adoption by firms and institutions.

In terms of AI-specific research, there is a more extensive body of literature than for GVC integration, encompassing both theoretical and empirical studies (Shih and Chang, 2009; Jiang, 2021; Tang et al., 2022; Tang, Li, et al., 2022; Zhao Rongying et al., 2019). This can be attributed to the availability of more recent data compared to the generally slower release of input-output tables, which underpin much of GVC research.

Going beyond this, it could also be speculated that knowledge flows through inter-firm collaboration, patents, and research networks are more immediate routes for AI knowledge diffusion than international trade, potentially bringing productivity benefits more rapidly than deep integration into global value chains.

### 2.3: Diffusion to LMICs: Benefits and Risks

As an increasing amount of research is done on AI economic impacts, worries of LMICs emerging as major losers in the AI race become more prevalent (Andersson Lipcsey, 2023; Artuc et al., 2022; Spence, 2022; Kouka and Magalles, 2022; Bughin et al., 2018; Brynjolfsson, Rock, Syverson, et al., 2017).

Some research presents an optimistic view, suggesting that an AI-aided shift towards more skill-intensive industries could enable LMICs to transform their economic structures and penetrate sectors previously inaccessible due to resource constraints (Kouka and Magalles, 2022). Additionally, LMICs highly integrated into GVCs may benefit from AI-powered reductions in trade frictions (Jayathilaka, 2022).

However, a significant share of the research points towards a bleak future for LMICs if the current trajectory persists. A common theme is the control over technology. Those who control the production and development of AI technologies, and adjacent technologies, are expected to be the big winners in the coming decades. As AI is integrated into economies and societies, those in control of the technology will wield substantial economic and political power over those who do not (Kouka and Magalles, 2022; Brynjolfsson, Rock, Syverson, et al., 2017). Other issues include the risk of on-shoring due to increased AI-enabled automation in manufacturing, leading to reduced foreign direct investment (FDI) (Artuc et al., 2022; Spence, 2022). Furthermore, AI may exacerbate disparities in LMICs by widening the gap between more advanced, internationally active firms that dominate exports and small, informal firms that employ a large share of low-skilled and manual labor (Artuc et al., 2022).

Given these potential impacts, it is increasingly important to understand the first step causing them: the diffusion of AI technologies to LMICs. This understanding is crucial to grasp the current situation, anticipate its evolution, and explore tools to mitigate severe impacts.

Turning to the first avenue of technology diffusion, GVC participation has been identified as a path for enhanced innovation in LMICs. These countries can leverage both GVC-provided knowledge and external resources to gain an edge in the global innovation race (Criscuolo and Timmis, 2017). Supporting this conclusion are findings that indicate a positive link between GVC participation and the adoption of Industry 4.0 technologies in developing countries (Delera et al., 2022).

Overall, participation in global value chains has shown a positive link with productivity growth for countries lagging behind the global productivity frontier, applicable across several economic sectors (Garcia, 2021; Rigo, 2020; Pahl and Timmer, 2019). On an economy-wide level, there is evidence that technologically lagging countries benefit comparatively more from foreign R&D spillovers that percolate through international trade (Foster-McGregor et al., 2016).

Regarding knowledge flows, technology clusters facilitating the diffusion of AI technologies have yet to form in many LMICs (Jiang et al., 2021). To close the gap with leaders in the field, focusing initially on disembodied forms of diffusion, primarily research, and subsequently on embodied forms, such as product engineering and development, seems preferable (Shih and Chang, 2009). Furthermore, countries with weak GVC participation can partially compensate for lost diffusion potential through active participation in research networks,  although it must be noted that findings in this case pertained to European countries and research networks (Daniela Di Cagno et al., 2014).

And while the importance of research collaboration for diffusion is sometimes downplayed, especially over large distances (Tang, Li, et al., 2022), it is also found that industrial distance has a positive relationship with AI international collaboration. In other words, countries with similar levels of industry involvement in AI research have fewer collaborative papers, as seen between the US and China (Tang, Li, et al., 2022). This presents an opportunity for LMICs to facilitate diffusion through research-induced knowledge flows.

Given the above findings, it may be tempting to conclude that indefinitely increasing the rate of AI diffusion to LMICs is the best solution to mitigate the effects of slow diffusion. However, the risks of rapid diffusion, particularly in developing nations, complicate this picture. Compared to the developed world, LMICs have more jobs likely to be replaced, rather than complemented, by AI technologies, potentially leading to more significant impacts. Conversely, in the developed world, with a higher proportion of high-skill-intensive occupations, a larger share of jobs is expected to be impacted by AI diffusion, providing a counterargument to the above statement (Pizzinelli, 2023). Still, if the larger percentage of complementary jobs is paired with efficient economic policies, this can lead to productivity gains and increased labor demand, outweighing job losses. Meanwhile, LMICs have significantly fewer resources for policies such as retraining programs or job guarantees. Therefore, the non-complementary nature of jobs is predicted to carry more weight for labor market outcomes, suggesting that in a scenario of rapid AI diffusion, labor markets in LMICs will suffer more damage than those in developed economies. This nevertheless remains a topic that warrants significantly more research.

Lastly, although slightly detached from the economic focus of this paper, another widely publicized risk of rapid AI diffusion to developing nations involves threats to democracy and fragile political systems. Developing nations typically have lower levels of democracy and fewer checks and balances against authoritarianism than developed countries. Given how AI can enable widespread surveillance and the dissemination of large-scale disinformation, there are concerns that rapid AI diffusion, combined with lagging AI governance policies, could push these countries towards more authoritarian and destabilized states (Naorobi and Paulo, 2024).

# 3. Methodology and Data

**3.1: Methodological framework**

Based on the findings of the literature review, a three-way approach to AI diffusion is suggested. These three avenues represent distinct avenues through which AI technologies disseminate to low- and middle-income countries (LMICs).

The first route is global value chains, which aligns closely with embodied diffusion. This route is expected to facilitate diffusion over the long term, as spillovers from AI-intensive sectors gradually reach non-intensive ones, and as demand for intermediate inputs, such as chips, increases to support AI advancements.

The second avenue is research networks, facilitating medium-term disembodied diffusion. Co-authored research projects provide a more rapid avenue for diffusion than global value chains due to the significant rigidity in the latter's structure. However, innovations in academia may take time to trickle down to firms, depending on the nature and discipline of study.

The third pathway involves direct knowledge transfers between firms participating in innovation clusters, presenting the fastest means of exchanging AI innovations and offering the highest likelihood of rapid implementation. This form of disembodied diffusion leverages the intense, direct collaboration within technological and innovation clusters.

While overlaps between these three avenues exist, categorizing them into distinct avenues enables easier empirical analysis and helps countries understand which pathways to focus on based on their economic strengths and weaknesses. This categorization also aids in determining an overall AI adoption strategy.

Patents, often used as a proxy for measuring diffusion have been chosen not to be included in the above framework, as they are seen more as outcomes of diffusion and proxies of technology intensity rather than direct avenues of diffusion. They are thus perhaps better suited to measure the next step in the process: productivity gains resulting from diffusion.

It is important to note that diffusion is understood as both an international and a national process. For diffusion to result in an increase in technology intensity, it is insufficient for only a few internationally exposed firms to gain access to new technologies. Smaller, less internationally active companies must also adopt and utilize the technology in order for its diffusion to permeate the economy and produce tangible impacts. Therefore, the analysis incorporates data on both international and domestic diffusion, where applicable. However, the primary focus remains on international diffusion, as it is assumed that in the case of advanced AI models, low- and middle-income countries (LMICs) with lower technology intensities will initially rely more on international diffusion rather than domestic innovation.

### 3.2: Choice of countries

Sixteen countries were chosen for analysis, determined by the limited number of overlaps between the datasets used. Unfortunately, these limitations excluded low-income countries from the investigation. According to the World Bank's 2020 classification, based on GNI per capita, all countries in the sample are lower-middle income (World Population Review, 2020). One exception is India, which was the 17th country to span all three datasets, was not included as it is a major AI innovator itself, and therefore not a country as dependent on inter-country AI diffusion as the rest of the sample. If anything, it can rather be seen as a significant source of diffusion  itself.

Additionally, Sweden, the United Kingdom, France, and Hungary were included to test the theory that the Global South is lagging behind the developed world in terms of AI diffusion. To understand the role of source countries in the diffusion of AI technologies to LMICs, the US and China were examined separately.

**3.3: Diffusion through global value chains**

For diffusion of AI technologies through global value chains, the OECD's 2023 TiVA indicators database (OECD, 2023) was used, which in turn are calculated based on the OECD's Inter Country Input Output database (ICIO). The ICIO tables are built up to the year of 2020.

More specifically, two TiVA indicators were chosen, both of which indicate depth of integration into global value chains. The first one being backward participation in global value chains (DEXFVAPSH), which captures foreign value added as a percentage of exports of the given country. Meanwhile, second one being forward participation (FEXDVAPSH), which stands for domestic value added in exports of other countries as a share of gross exports of the chosen LMIC. Essentially, the first metric indicates how much technology the LMIC is receiving from abroad, while the second metric reflects how much of its domestic innovations the nation is contributing to international GVCs.

One advantage of these measures is that they capture multi-level, indirect links, not just direct trade flows, providing a more comprehensive picture of trade relations (Andersson Lipcsey, 2021, pp. 38-45). Additionally, they allow for the examination of the intensity of trade relations with specific countries, particularly the USA and China, as selected for this study.

Data was collected only for the most recent year available, as trade intensity is viewed more as a predictor of future diffusion rather than an ongoing process. Although supply and use tables (SUTs) from the years 2021-2023 might capture novel trends attributable to AI diffusion, it will take a few years before they are incorporated into the latest version of the OECD ICIOs.

### 3.4: Diffusion through research networks

To quantify diffusion through research networks, the OECD AI Policy Observatory's dataset, sourced from OpenAlex (OECD AI Policy Observatory, 2024), is utilized for national and international collaborations in AI scientific publications. This dataset captures both domestic and international diffusion through co-publication nodes between and within all countries in the study.

Data spans from 2010 to 2023, providing insights into trends and changes over time. To mitigate the impact of large-scale research projects that could skew findings, averages are calculated for various years. These findings are then normalized with population data from the UN's 2023 World Population Prospects Database (United Nations Population Division, 2023) to ensure comparability across countries.

By averaging and normalizing the data, this approach allows for a balanced comparison of AI research activity and collaboration intensity among countries, highlighting the extent and effectiveness of AI diffusion through research networks.

### 3.5: Diffusion through knowledge transfers

For direct knowledge transfers, the OECD AI Policy Observatory's dataset on knowledge transfers based on Stack Overflow requests data is employed (OECD AI Policy Observatory, 2023). Stack Overflow, a question-and-answer website for computer programmers, provides insights into requests made within AI technologies, capturing both domestic and international flows.

However, there are some limitations to this dataset. It does not include data for Ivory Coast and Laos, and there is an unexplained decrease in fulfilled requests between 2022 and 2023. This decline could be due to poor data quality or the rise of ChatGPT as the primary source for programming-related questions. Consequently, unlike the diffusion through research where 2021-2023 averages are used, for knowledge transfers, 2019-2021 averages are employed to ensure more reliable analysis.

### 3.6: Comparisons with developed nations

Quantifying the rate of AI diffusion to developing countries is valuable, but context is crucial to interpret these results meaningfully. For instance, if a country ranks first in AI publications per million inhabitants among developing nations, it is essential to compare the dynamics of this to developed nations to understand if the "gap" between developed and developing countries is narrowing or persisting. How does this gap look today, and has it changed over recent years?

To address these questions, four developed economies are included for comparison: the United Kingdom, Sweden, France, and Hungary, representing various levels of AI diffusion in the developed world. Notably, comparisons for GVC integration are not made here, as this metric indicates diffusion potential rather than current diffusion. Additionally, trend comparisons for GVC integration would be less meaningful due to the rigid nature of input-output tables. Therefore, only AI co-publications and knowledge flows are considered.

Both domestic and international diffusion are included in this analysis, as the goal is to understand the current and evolving gap between developed and developing nations. Furthermore, the analysis seeks to assess the development of domestic knowledge clusters in the countries investigated. This metric is significant because it indicates whether a country has well-developed domestic knowledge clusters to facilitate the proliferation of AI technologies domestically. It also provides insight into how vulnerable these countries might be to global protectionist trends that could limit international diffusion.

# 4. Findings

## 4.1: Inter-country AI diffusion through global value chains

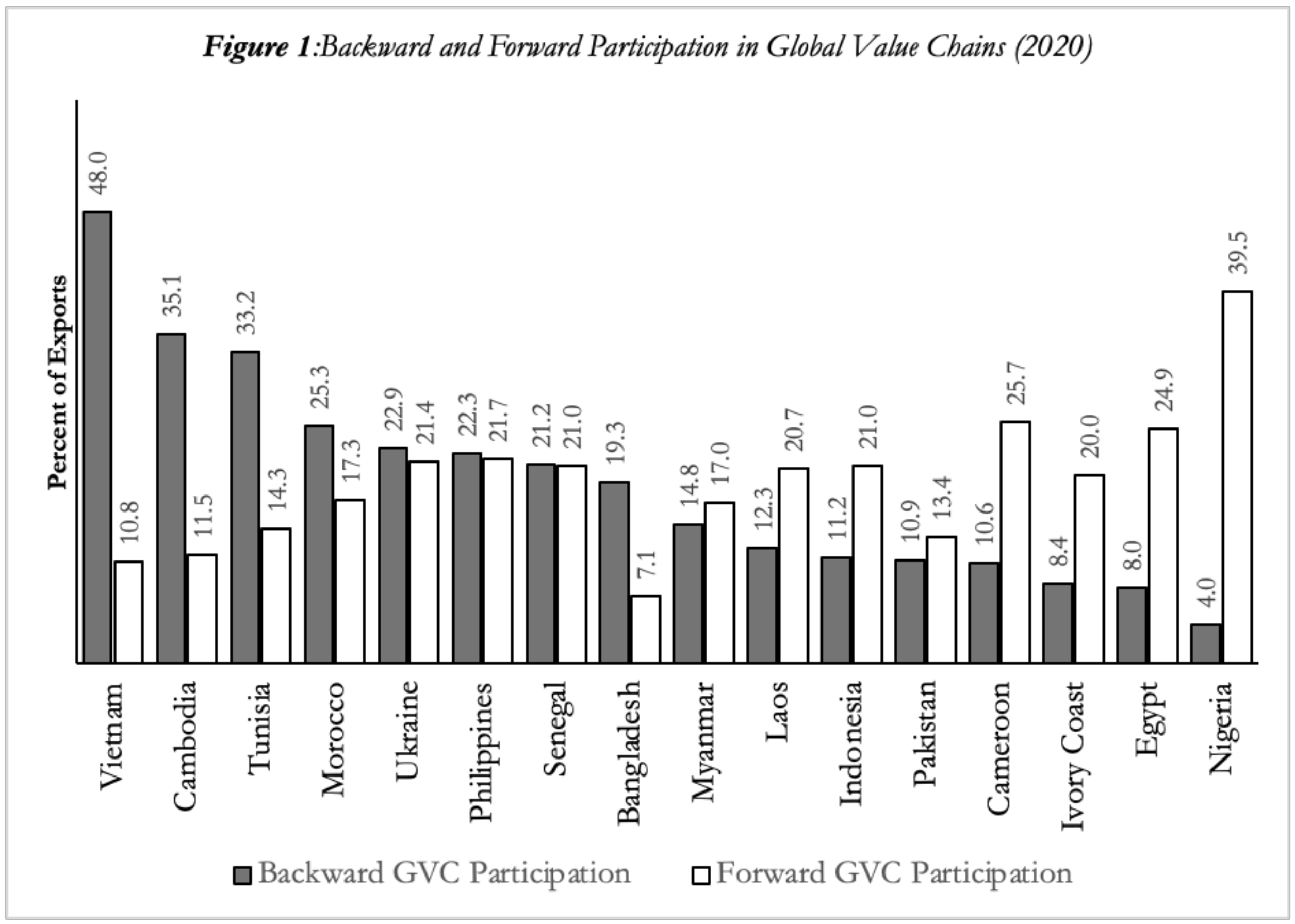

***Figure 1***:*Backward and Forward Participation in Global Value Chains (2020)*

As seen in **Figure 1**, GVC integration of the countries analyzed reveals a non-complementary relationship between backward and forward integration. For instance, Vietnam, which has the highest percentage (48 percent) of its exports containing foreign value added, exhibits minimal domestic value added in foreign exports. This discrepancy is likely due to the substantial inflow of intermediate goods into Vietnam for manufacturing, which are then re-exported. Conversely, Nigeria has a high level of raw materials that are extracted, processed, and exported for incorporation into the manufacturing of various technology products abroad, such as in China, which then exports the finished products globally. Countries like Ukraine and the Philippines show a greater balance between backward and forward integration.

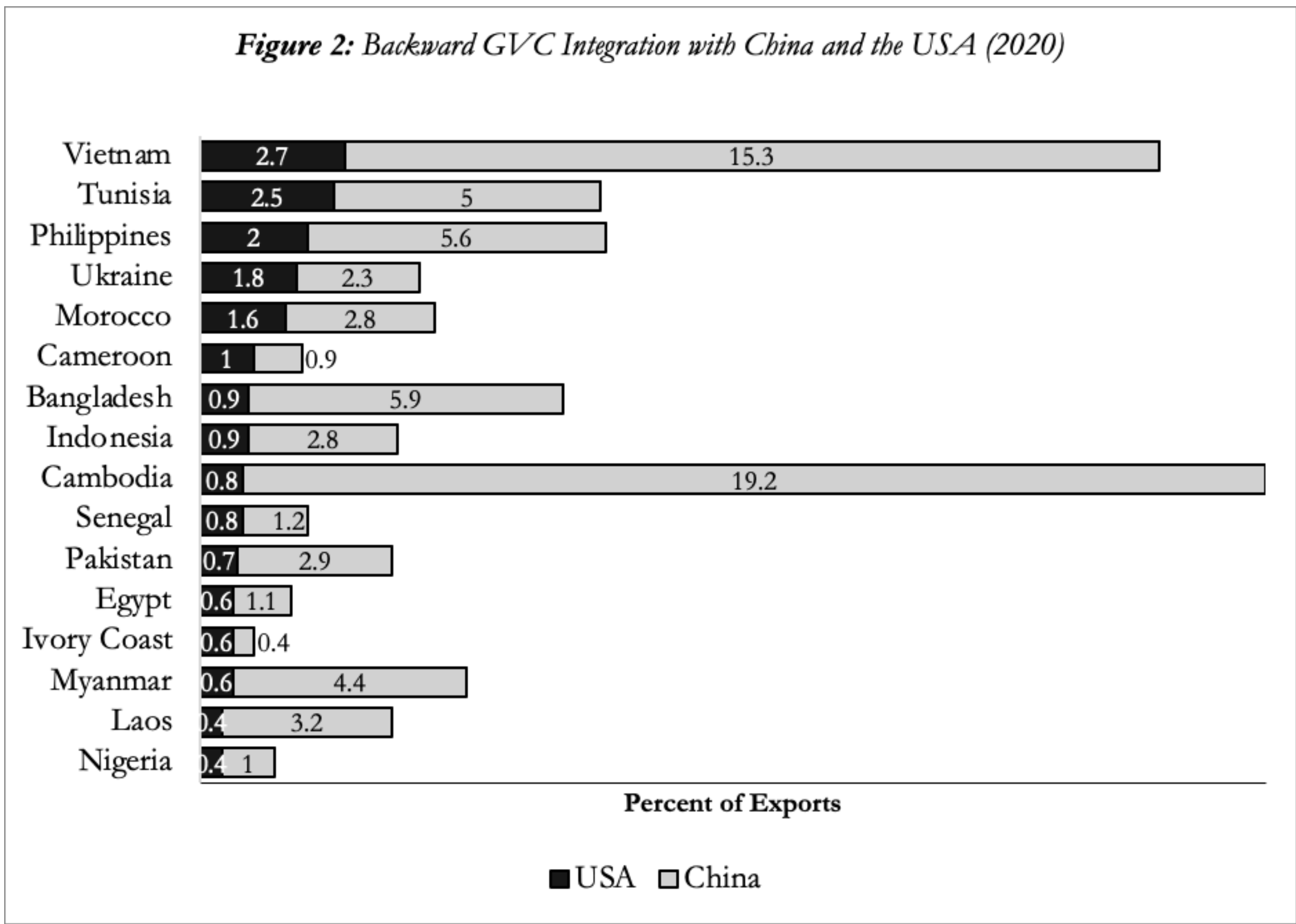

***Figure 2:*** *Backward GVC Integration with China and the USA (2020)*

As seen in Figure 2, Vietnam and Cambodia stand out regarding the inflow of Chinese and American value added, both with a significant amount of Chinese value added (15 and 19 percent, respectively) in their exports. Conversely, a surprisingly low percentage of their exports contain American value added. This disparity might be due to the USA acting primarily as a final destination for exports, whereas China is more engaged with intermediate goods. For instance, a significant portion of rare earth minerals used for chips are extracted in the Global South and almost entirely exported to China. China processes these minerals and then re-exports them, often to the US.

### 4.2: Inter-country AI diffusion through research networks

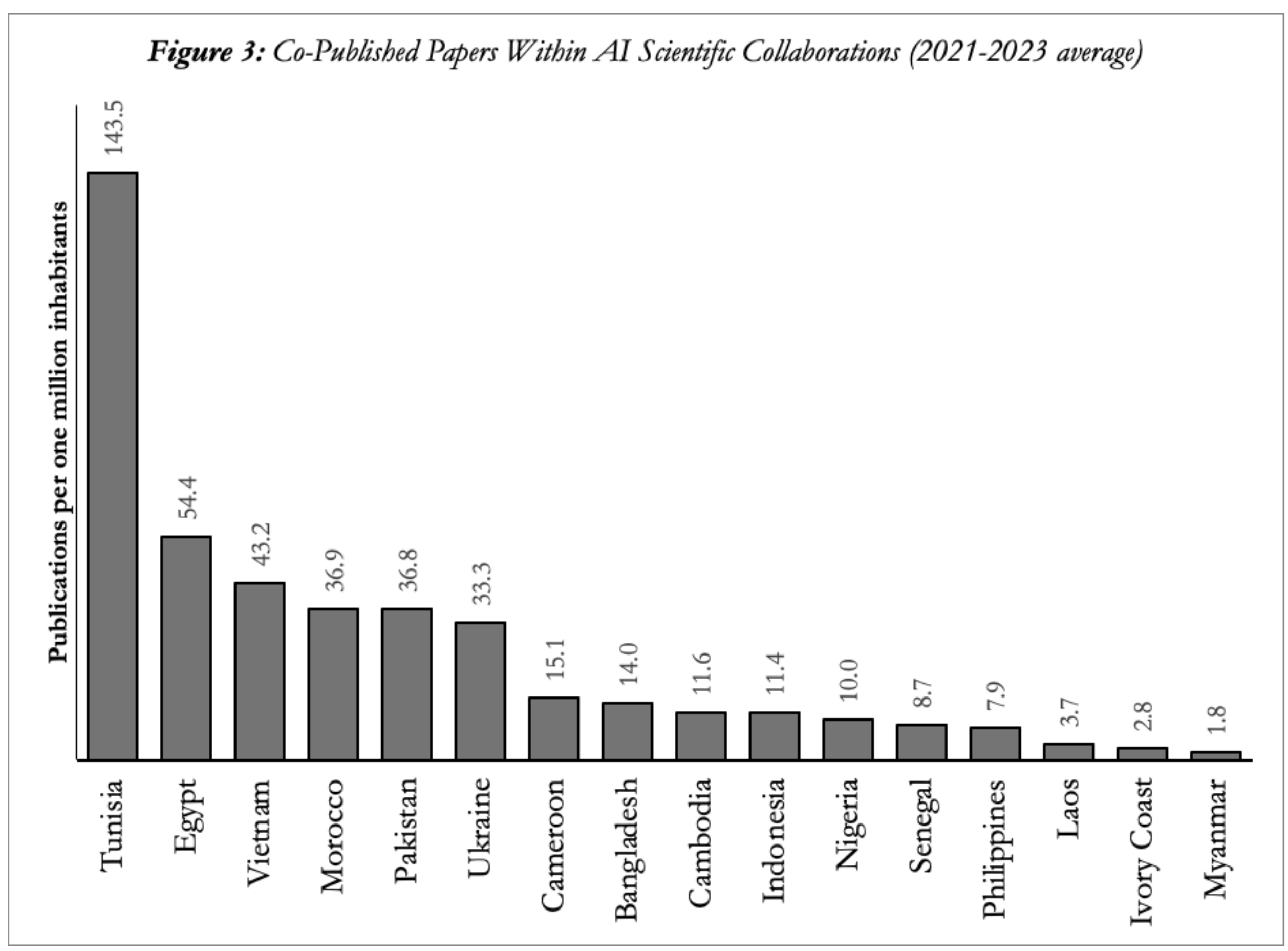

***Figure 3:*** *Co-Published Papers Within AI Scientific Collaborations (2021-2023 average)*

As can be seen in Figure 3, Tunisia stands out in terms of diffusion through research networks, with 144 international AI co-publications per one million inhabitants. Egypt and Vietnam follow, showing substantial engagement in international AI research collaborations. On the other hand, Myanmar, the Ivory Coast, and Laos have the fewest international AI co-publications, ranging between 1.8 and 3.7 per one million inhabitants.

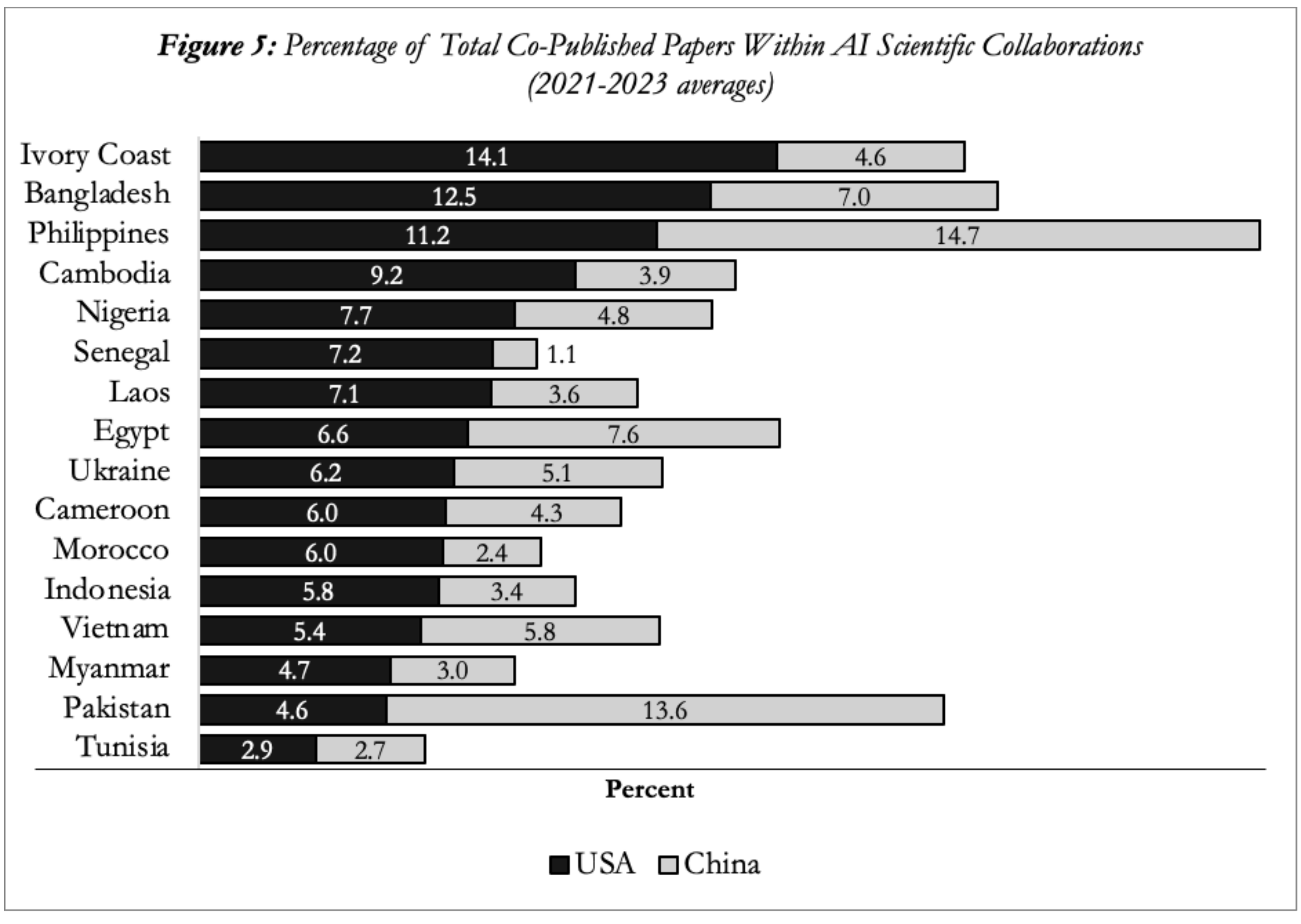

***Figure 5:*** *Percentage of Total Co-Published Papers Within AI Scientific Collaborations (2021-2023 averages)*

For research collaborations, integration with the USA appears to be more significant than through global value chains. As noted earlier, research collaborations are considered a more immediate source of cross-national AI diffusion. Hence, in the shorter term, the results presented in Figure 5 indicate a greater reliance (on average 7 percent) on the USA for receiving such technologies. Nevertheless, China is not far behind, contributing to an average of 5.5 percent of research papers. Additionally, China plays a significant role in collaborative research with the Philippines and Pakistan, co-authoring 14.7 and 13.6 percent of AI research papers, respectively.

**4.3: Inter-country AI diffusion through knowledge transfers**

As for AI knowledge transfers, as shown in Figure 6, Tunisia ranks at the top, with 11.5 AI Stack Overflow requests per one million inhabitants on average between 2019 and 2021, followed by Pakistan (7.3) and Ukraine (6.1). In contrast, Senegal had zero AI Stack Overflow requests, and both Cameroon and Myanmar sent out fewer than 0.5 requests per one million inhabitants. Additionally, as previously noted, data was unavailable for the Ivory Coast and Laos.

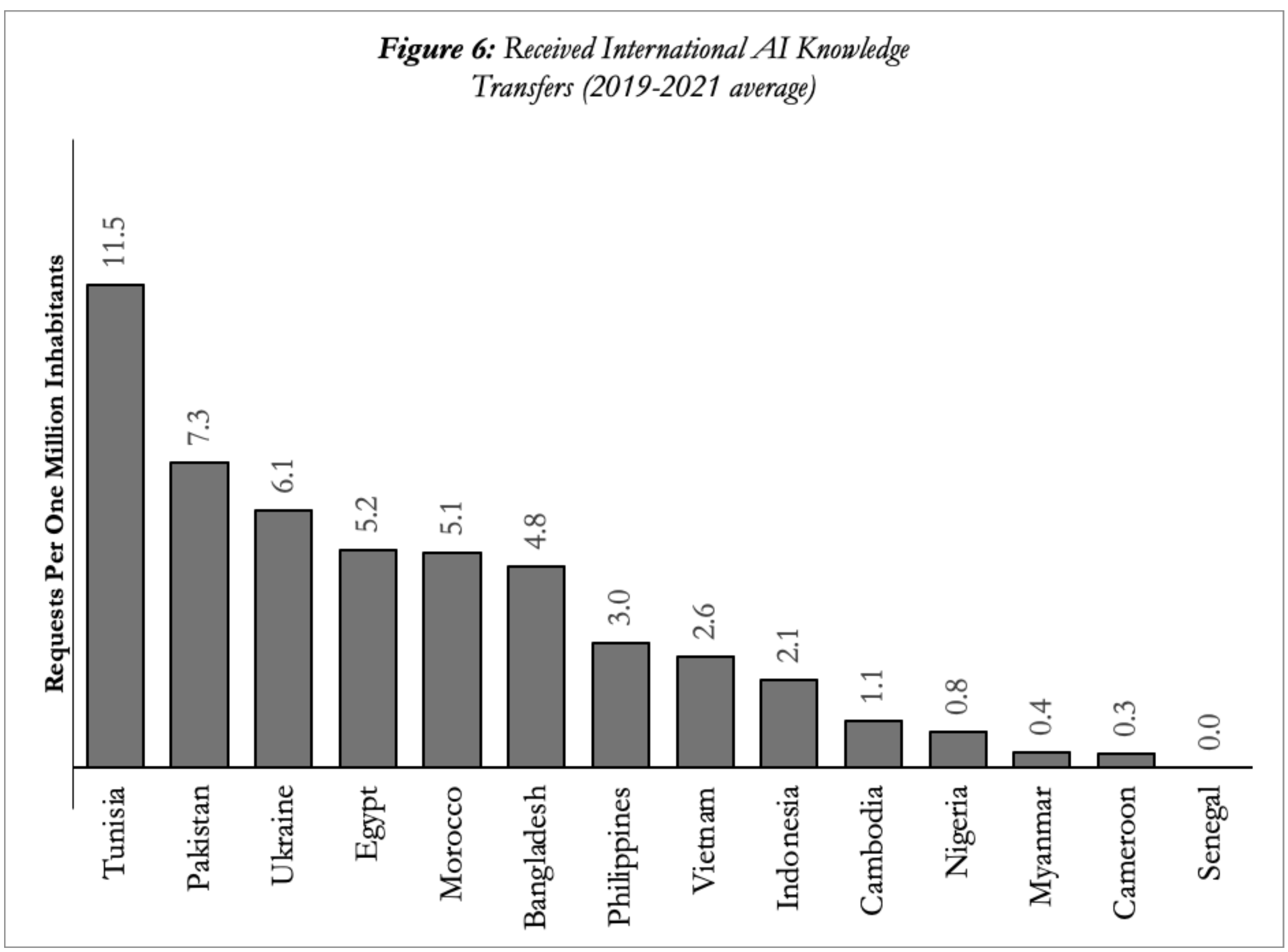

***Figure 6:*** *Received International AI Knowledge Transfers (2019-2021 average)*

Reliance on the USA is significant in the domain of AI knowledge transfers, as observable in Figure 7. For example, Nigeria received answers to 27.7 percent of its AI-related questions from the USA. On average, 17.3 percent of AI-related requests were answered from the USA, while only 0.97 percent were answered from China. This discrepancy may be due to a combination of factors, including a relative lack of expertise in the field, language barriers, and lesser popularity of the platform in China.

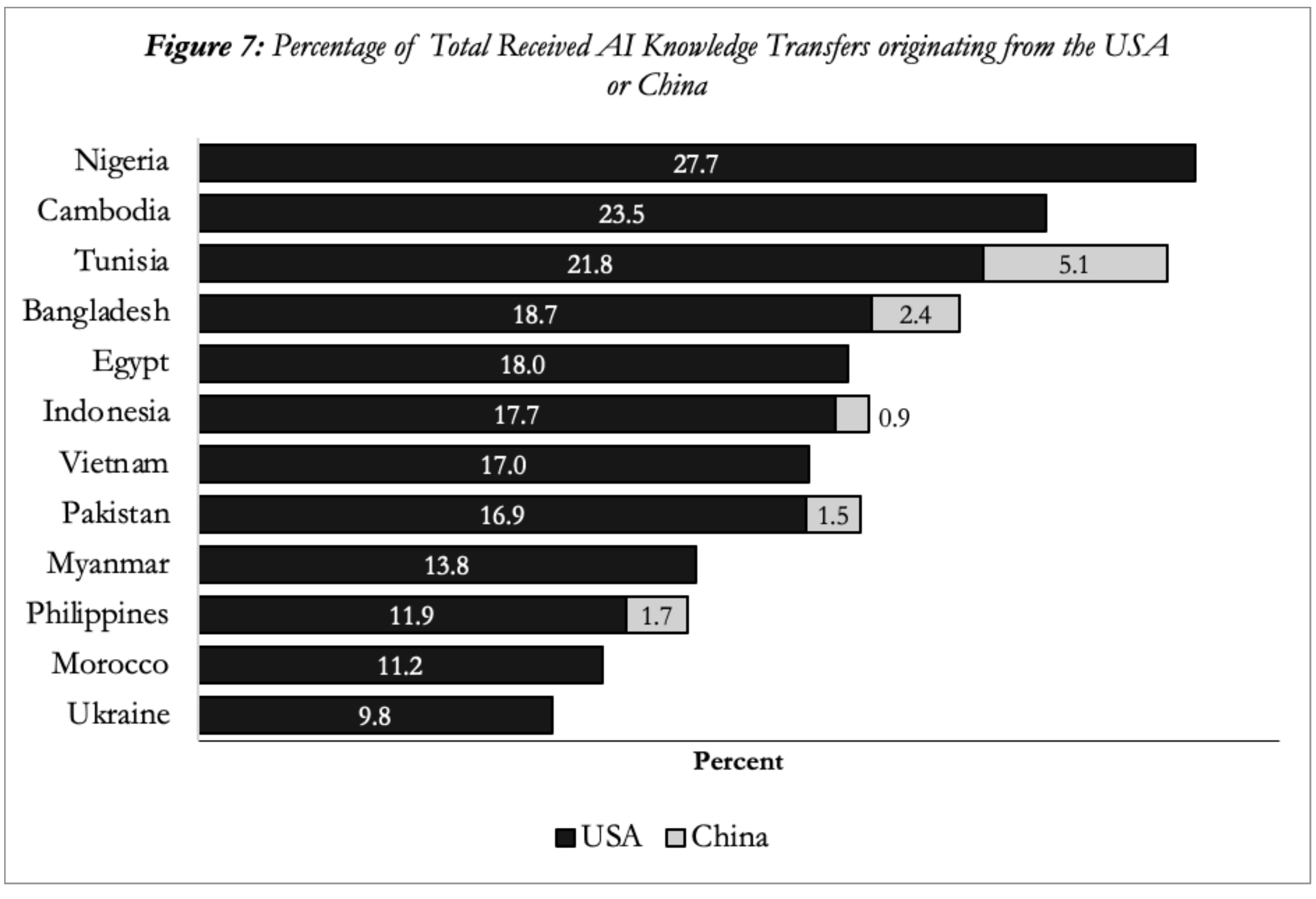

***Figure 7:*** *Percentage of Total Received AI Knowledge Transfers originating from the USA or China*

### 4.4: Total diffusion through research compared with the developed world

When compared to the four developed countries, the total (domestic and international) diffusion through research reveals that the examined low-middle income countries (LMICs) lag unexpectedly behind. As seen in Figure 8, Tunisia, the best-performing LMIC, had around one-fifth of the publications per one million inhabitants compared to Sweden in the years 2021-2023.

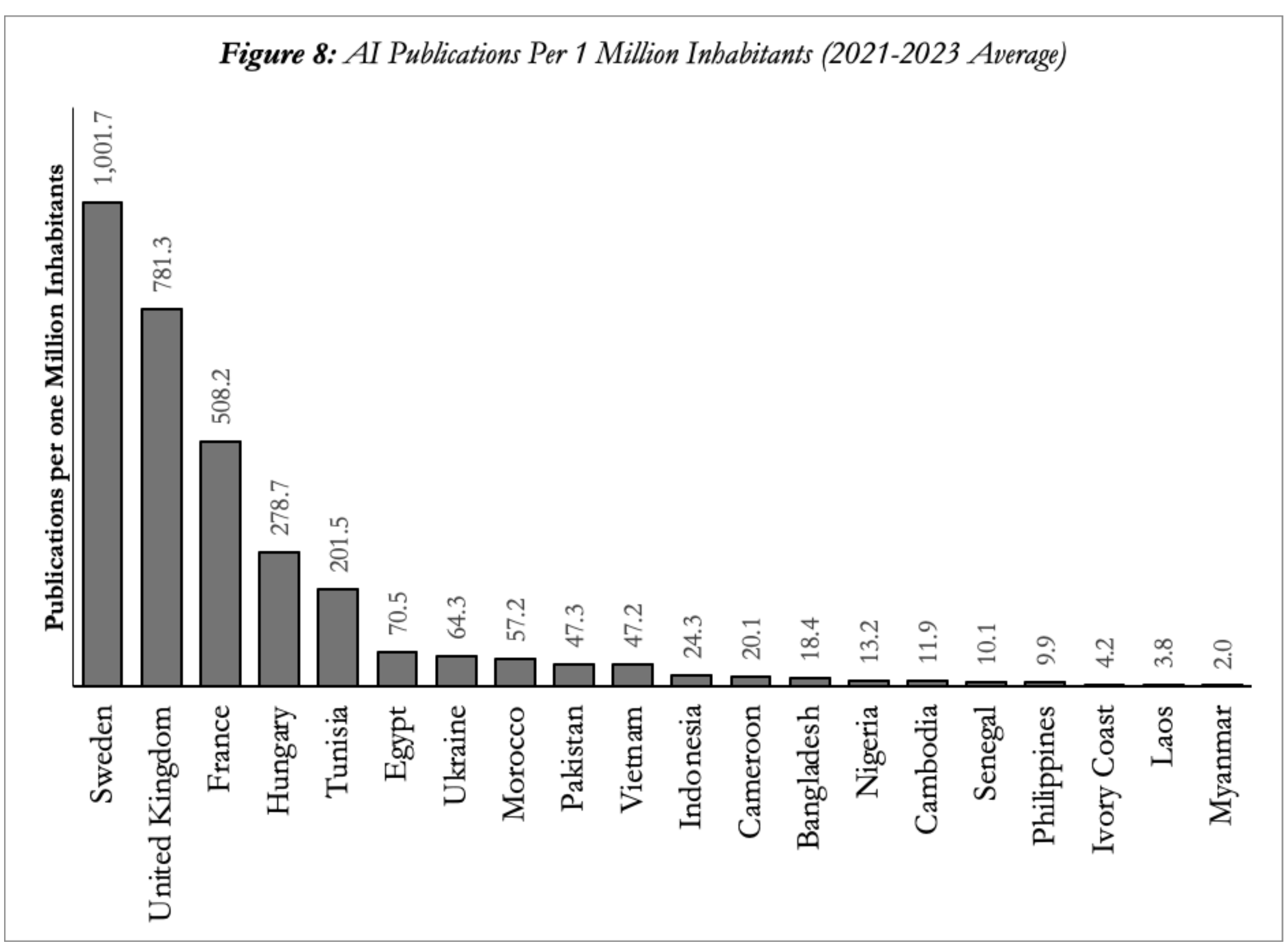

***Figure 8:*** *AI Publications Per 1 Million Inhabitants (2021-2023 Average)*

It may be argued that this does not provide the full picture, as trends in diffusion are equally important to understand whether the "gap" is closing or widening. Figure 9 below depicts a comparison between 2018-2020 and 2021-2023 averages, essentially capturing how much diffusion through research networks has accelerated. Overall, it is clear that diffusion has accelerated more in the LMICs than in the four developed nations, with an average increase of 60.1 percent compared to 25.4 percent in the latter group.

Furthermore, it can be observed that Bangladesh saw the biggest increase in AI publications per one million inhabitants, with a rise of 128.5 percent, while Myanmar was the only country in the sample that saw a decrease, with a drop of 9.7 percent.

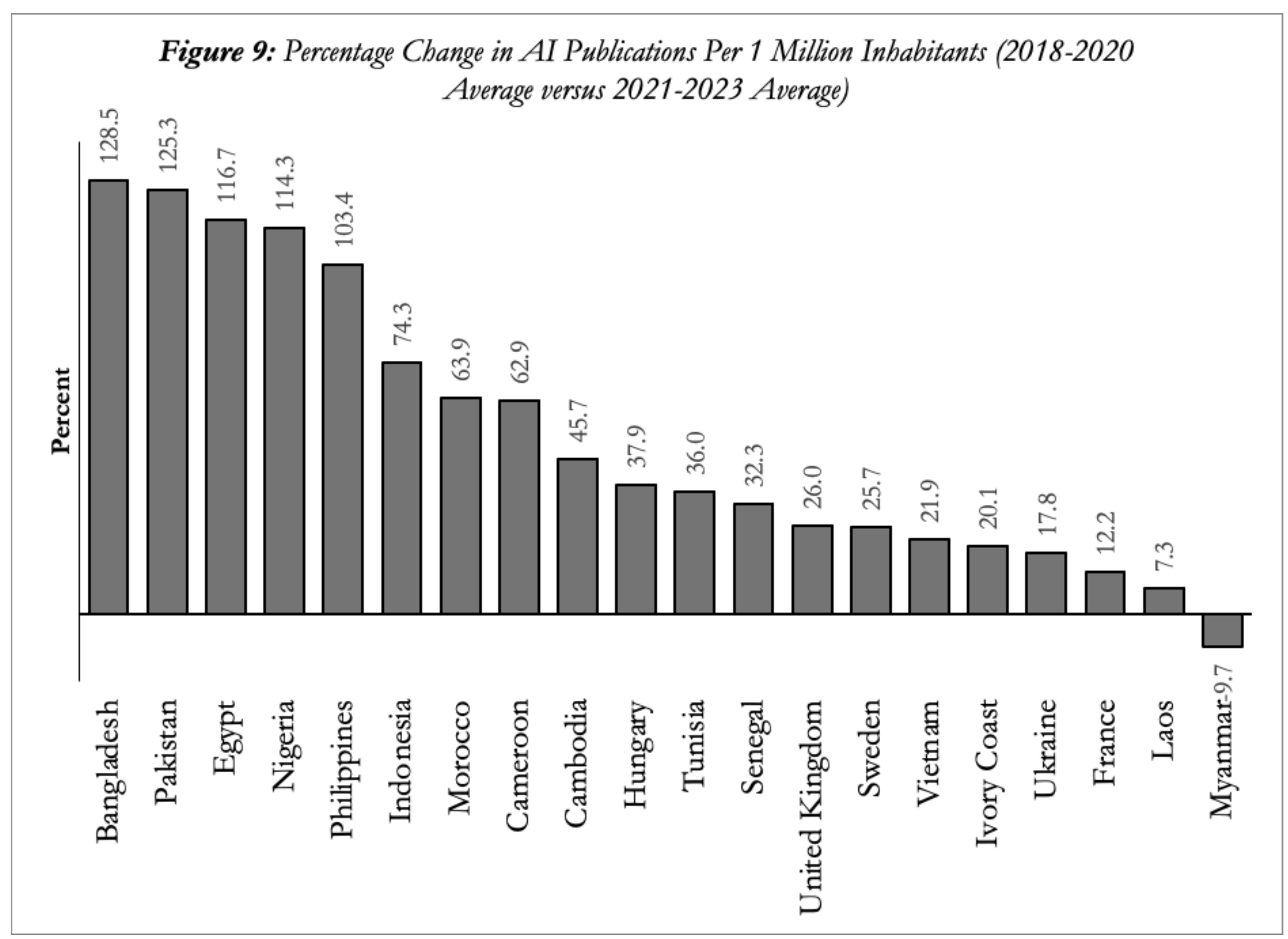

***Figure 9:*** *Percentage Change in AI Publications Per 1 Million Inhabitants (2018-2020 Average versus 2021-2023 Average)*

Another aspect to consider is how much of the diffusion is domestic. As seen in Figure 10 below, 53.1 percent of AI research collaborations in Indonesia were domestic, while in Cambodia, this figure was only 3 percent. Among the developed nations, France had a high percentage of domestic collaborations at 40 percent, whereas Sweden had the least, with only 17.1 percent.

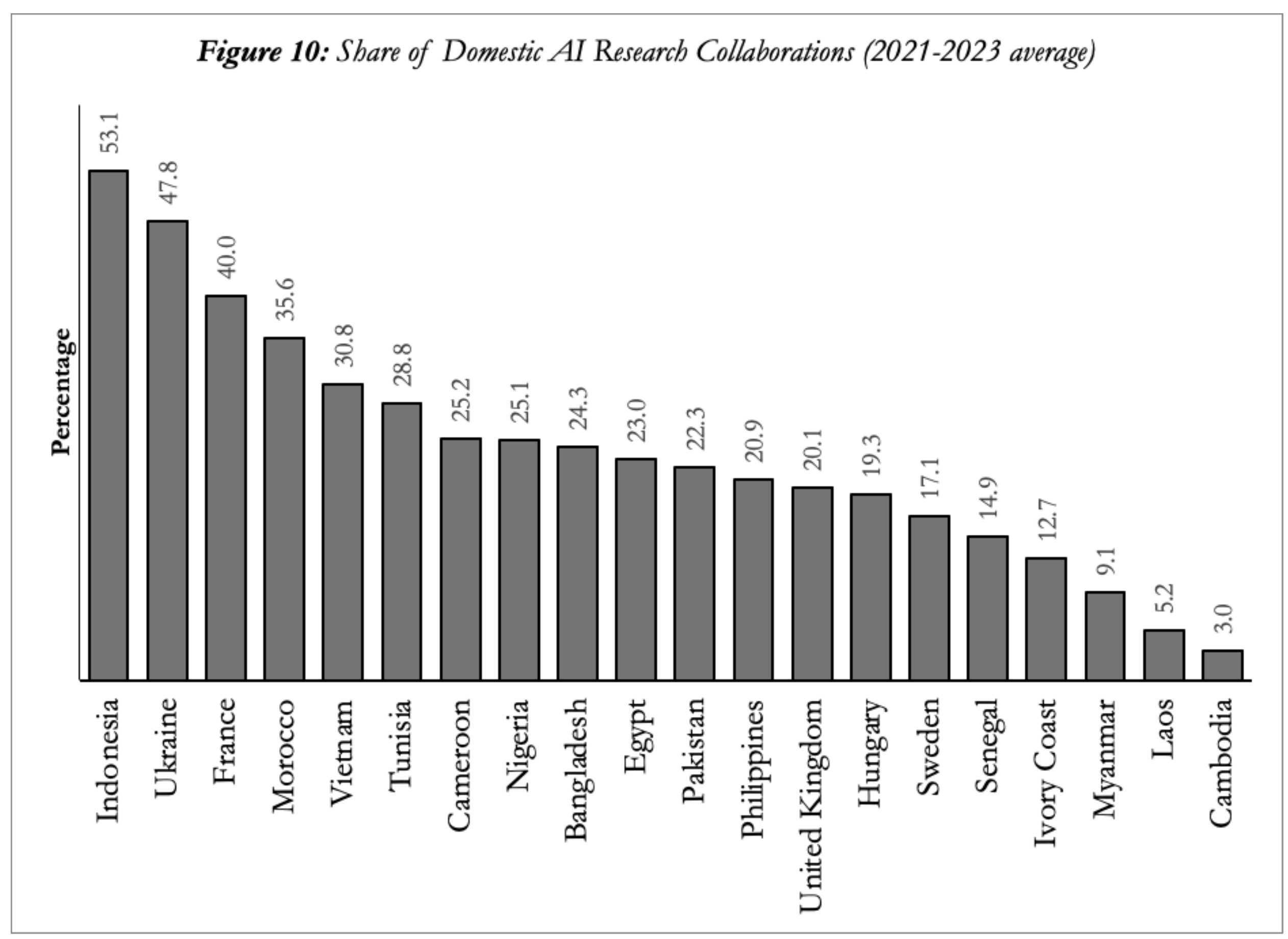

***Figure 10:*** *Share of Domestic AI Research Collaborations (2021-2023 average)*

In terms of trends, Figure 11 below shows that most economies experienced a decrease in domestic collaborations, with an average drop of 4.9 percent. However, Indonesia, Ukraine, France, Morocco, and Vietnam are outliers, as they saw an increase in domestic collaborations between 2018-2020 and 2021-2023. Cambodia experienced the most significant decline in domestic collaboration, with a drop of 25.2 percent.

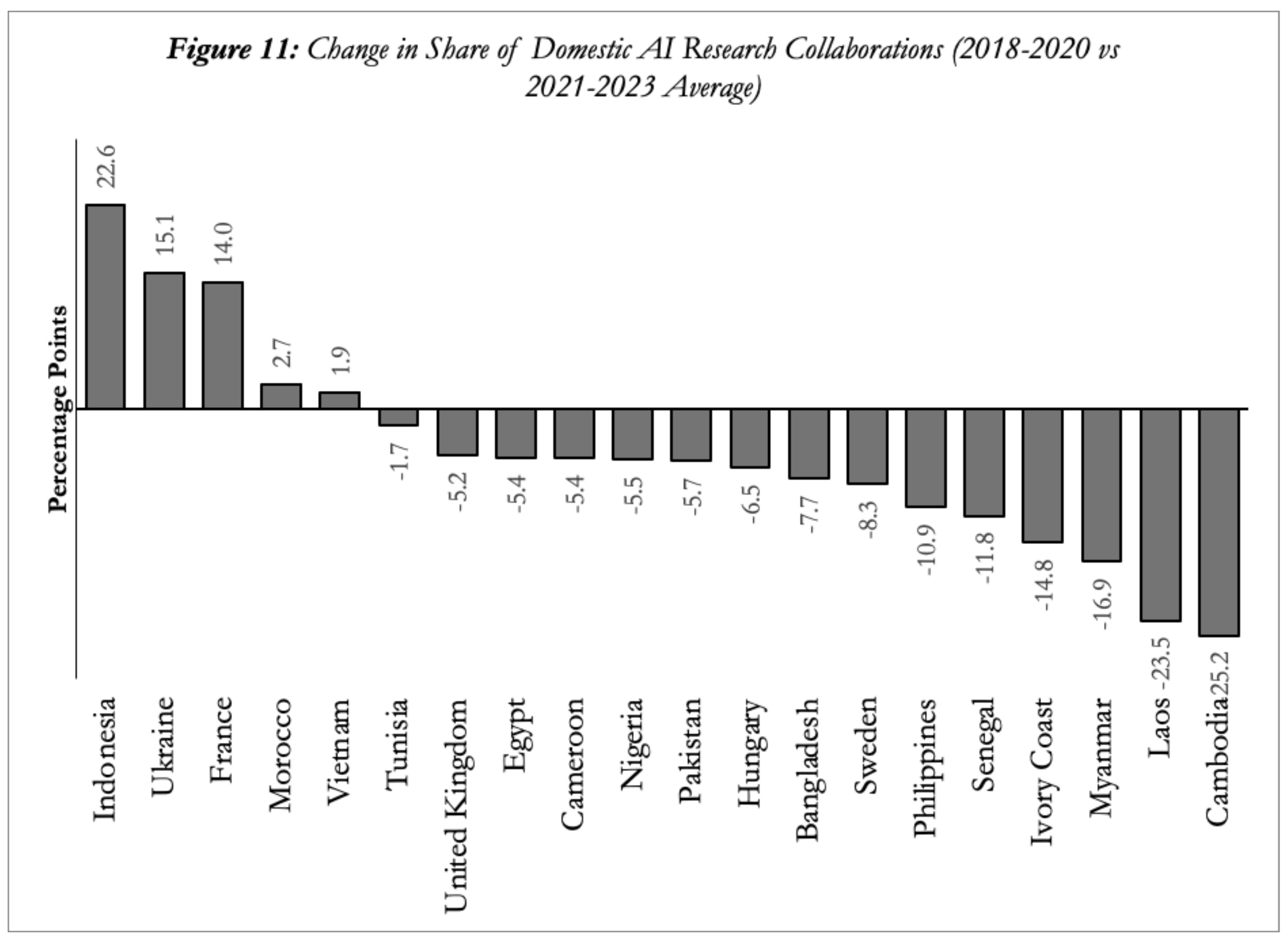

***Figure 11:*** *Change in Share of Domestic AI Research Collaborations (2018-2020 vs 2021-2023 Average)*

### 4.5: Total diffusion through knowledge transfers compared to developed nations

In terms of AI-related knowledge flows (both domestic and international), as observable in Figure 12, the four developed countries unsurprisingly take the lead, with the United Kingdom in first place, with 71.9 Stack Overflow requests per one million inhabitants. This figure is over 350 times higher than that seen in Senegal. The average number of Stack Overflow requests per one million inhabitants for the developed nations stands at 48.7, compared to 4.4 in the LMICs.

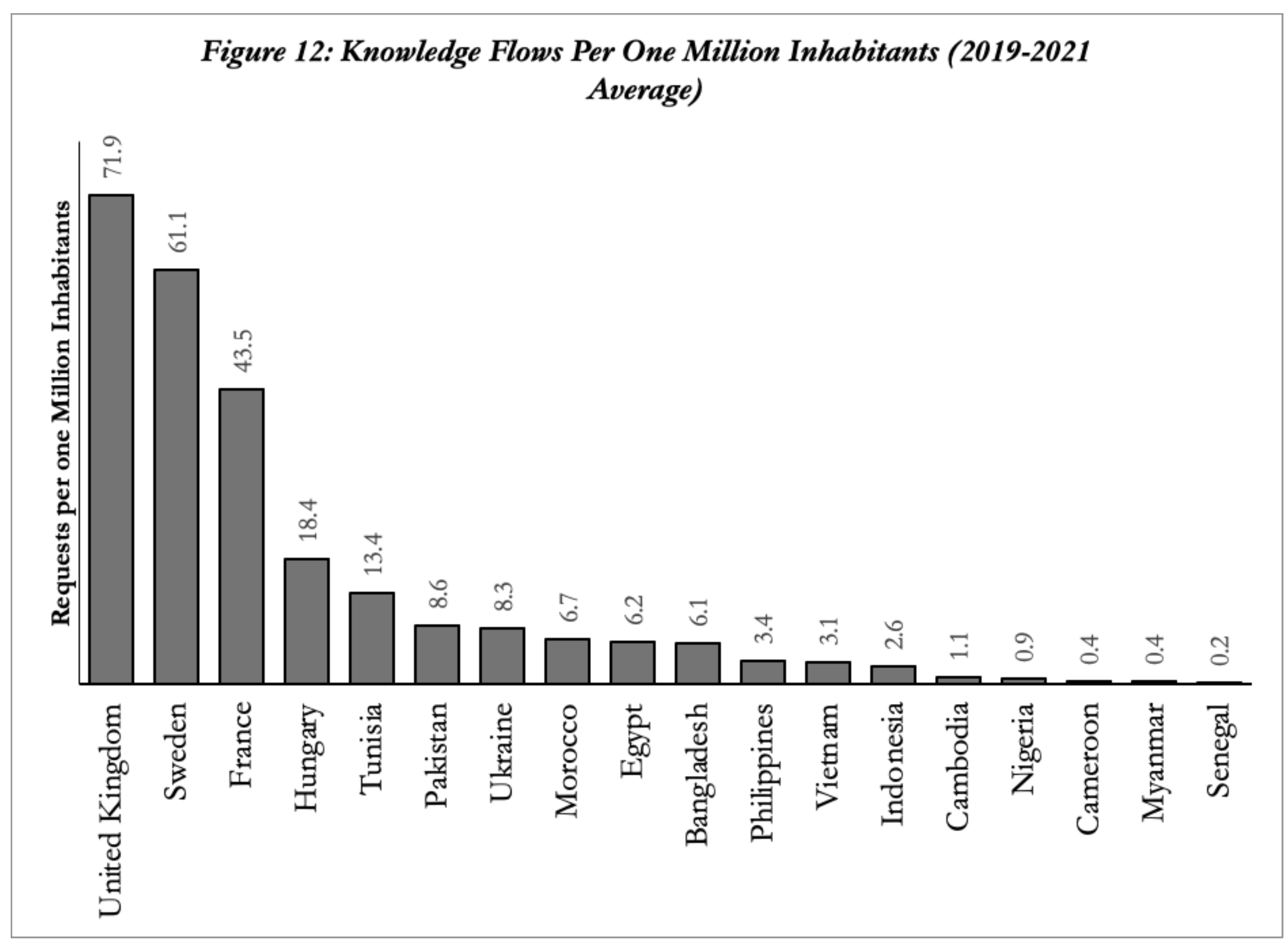

***Figure 12: Knowledge Flows Per One Million Inhabitants (2019-2021 Average)***

As seen in Figure 13 below, trends in diffusion through Stack Overflow requests are somewhat challenging to interpret due to the extreme variations. Myanmar, for instance, saw an increase of 625 percent, while the developed economies remained mostly flat, and Senegal experienced a decrease of 79.2 percent. These extremes are due to the very small number of requests in countries like Myanmar, which had 8 requests in 2018, 26 in 2019, and 32 in 2020. Nevertheless, excluding Myanmar from the calculation, the LMICs saw an average increase of 42.7 percent in requests, compared to a decrease of 6.5 percent for the four developed economies.

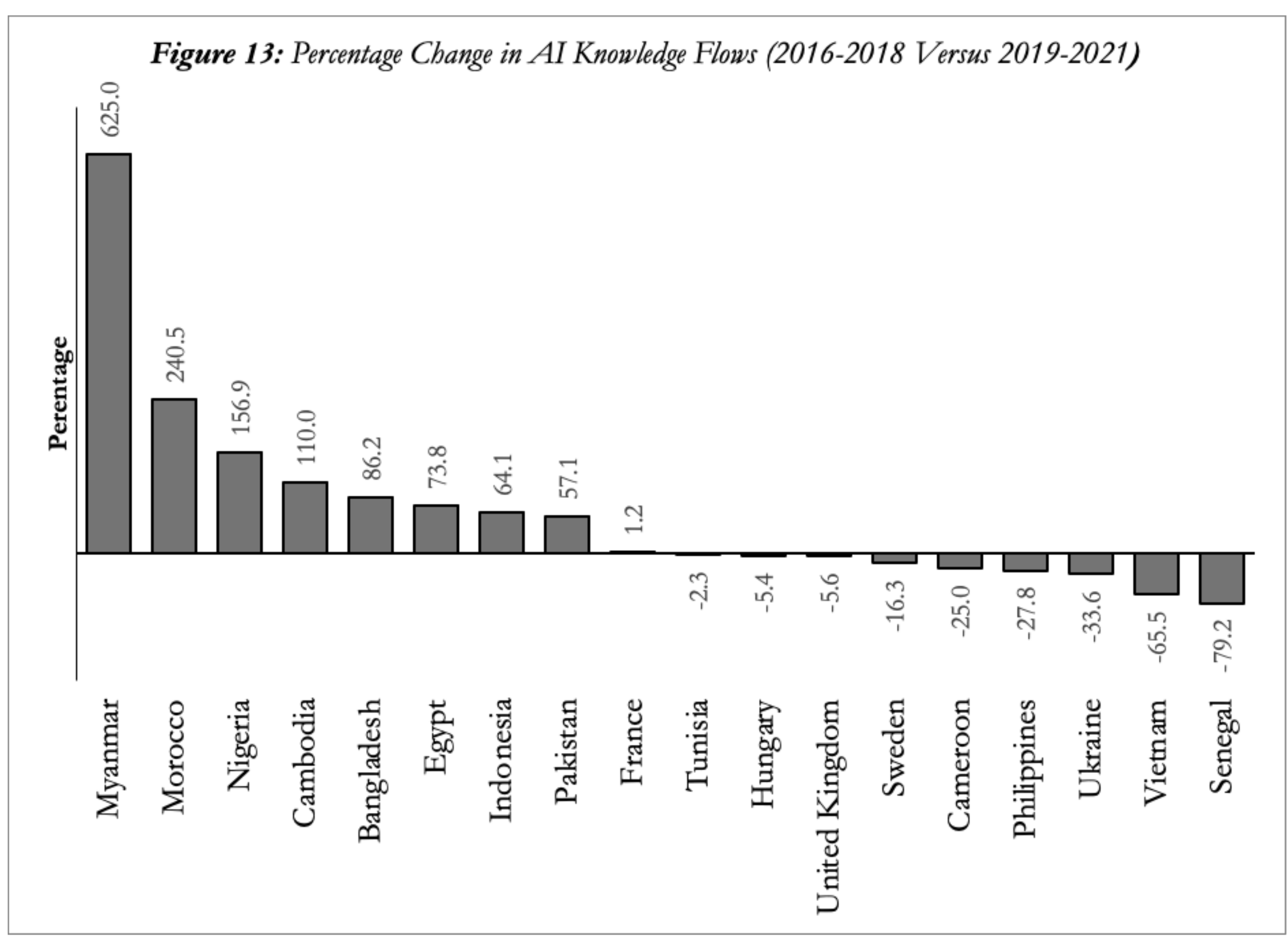

***Figure 13:*** *Percentage Change in AI Knowledge Flows (2016-2018 Versus 2019-2021)*

In terms of domestic knowledge transfers, France, the United Kingdom, and Ukraine stand out, with domestic Stack Overflow requests constituting 30 percent, 26.6 percent, and 26 percent of their total requests, respectively. At the bottom of the list are Cambodia, Senegal, and Myanmar, each with zero domestic requests, which is reflective of their already low number of overall requests.

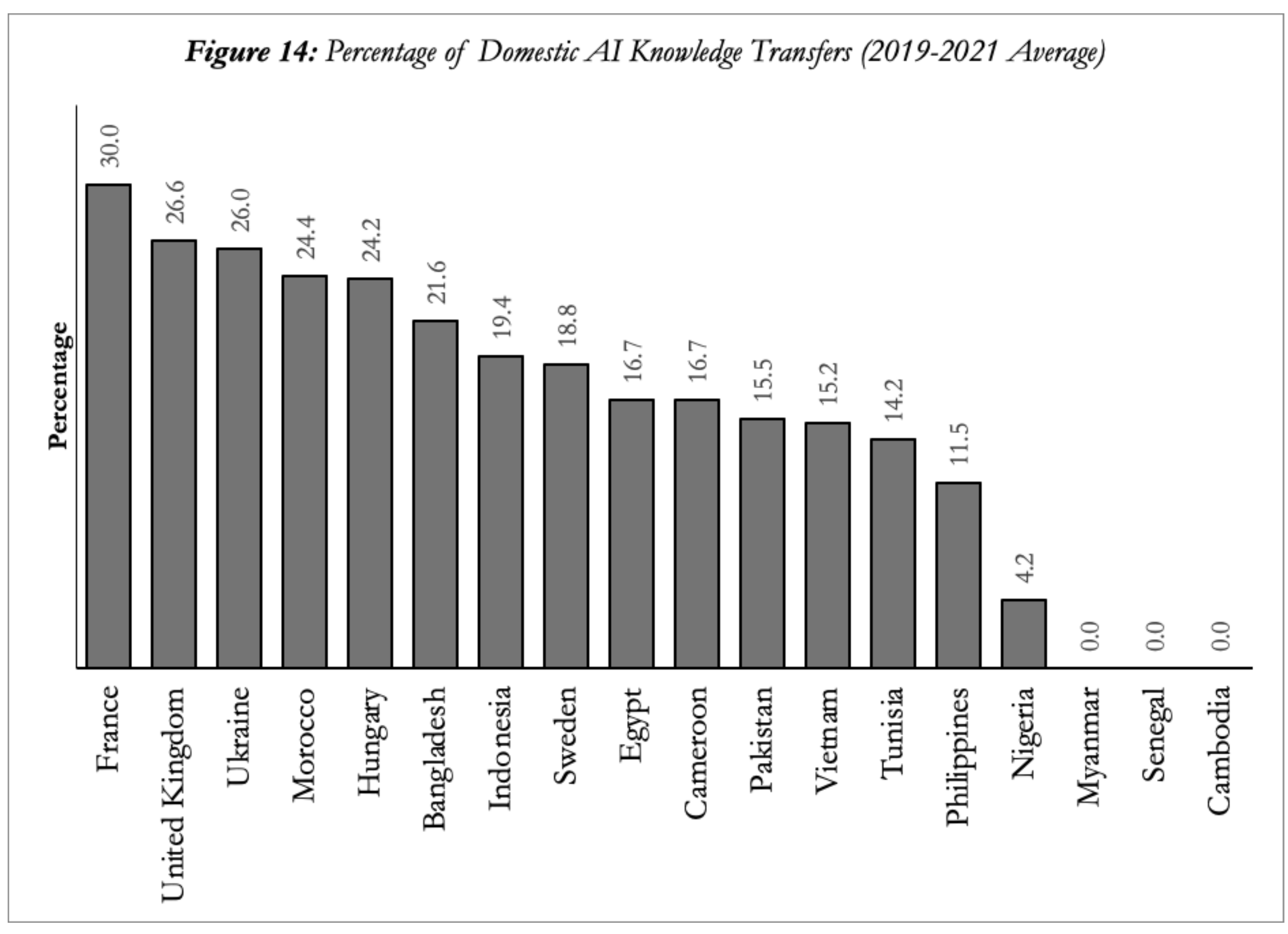

***Figure 14:*** *Percentage of Domestic AI Knowledge Transfers (2019-2021 Average)*

As for changes over the years, as seen in Figure 15 below, France and Indonesia once again see increases in domestic knowledge transfers, along with a few other countries. Overall, the share of domestic knowledge transfers grew by 8 percent, compared to a decrease of 4.9 percent for research networks. For the LMICs, there was an increase of 8.3 percent, while the developed economies saw an increase of 6.9 percent.

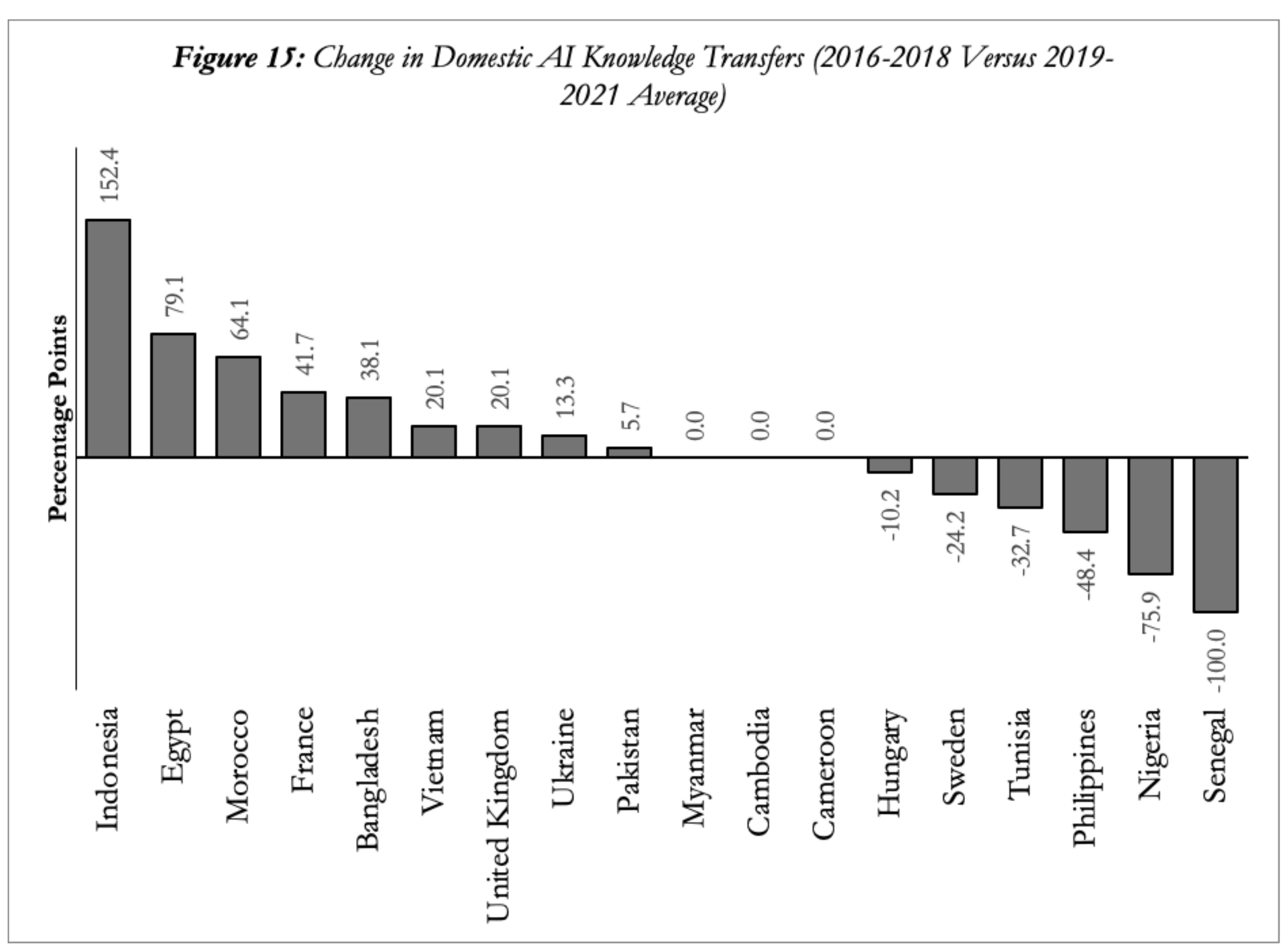

***Figure 15:*** *Change in Domestic AI Knowledge Transfers (2016-2018 Versus 2019-2021 Average)*

# 5. Discussion and Limitations

The goal of this paper was to provide a novel framework for understanding the avenues of international AI diffusion and to make an initial empirical contribution to the literature on diffusion to low- and middle-income countries (LMICs). While this objective has been achieved, it was evident from the outset that the paper would likely raise more questions than it could answer. Consequently, the discussion of results, limitations, and directions for further research is structured around key questions that emerge from the study.

**5.1: What data was omitted?**

The selection of LMICs for analysis in this study was determined by the overlaps in available datasets, which posed a limitation by precluding a more deliberate selection process. As previously discussed, no low-income countries were included due to data scarcity; consequently, lower-middle-income economies were analyzed instead. Ideally, the study would have featured a balanced representation across low, lower-middle, and middle-income groups to facilitate more specific and less generalized conclusions. Additionally, the inclusion of only four developed economies was due to time constraints, and a broader inclusion of first-world countries would have yielded more precise comparisons between LMICs and developed nations.

In terms of metrics for measuring diffusion, the inclusion of additional indicators such as patent data could have been advantageous. Although patent data has been featured in some studies within the literature review, it was excluded here due to capacity constraints and its characterization as an outcome of diffusion rather than a facilitator. Similarly, FDI data was omitted, despite its relevance. FDI, though interrelated with international trade and knowledge flows, represents another pathway for AI diffusion.

Furthermore, the study utilized data from 2019-2021 for knowledge flows, omitting 2022-2023 due to a notable decline in fulfilled Stack Overflow requests during the latter period. This decline may result from poor data quality or the emergence of AI tools like ChatGPT as replacements for human-to-human queries. Were data on GPT interactions available, it would likely constitute a substantial portion of recent knowledge transfers. However, this raises complex questions about the origins of these transfers, whether they are predominantly from the USA, where OpenAI is based, or from the global contributors to the model's training data. Such inquiries are expected to be the focus of future research, but in the short term, they underscore the challenges in interpreting ChatGPT's role in international knowledge diffusion.

**5.2: What are the limitations of the methodology employed?**

One limitation of the methods utilized, which will be discussed in more detail later, is the uncertainty regarding how quickly the three avenues of diffusion lead to increases in technology intensity. Additionally, we lack information about the baseline technology intensities of the countries included in the study. The methods employed can indicate which countries have experienced, are experiencing, and are expected to experience higher rates of diffusion. However, they do not provide precise information on the extent or timing of these increases.

This is particularly relevant for GVC integration, which not only lacks temporal perspectives, but also fails to capture sectoral heterogeneity in diffusion. For instance, if we observe that country X overall has a low backwards integration with technology source country Y, at the absence of additional information, we may presume that country X will not see major diffusion from country Y through the pathway of international trade. Yet, upon deeper analysis, we may find that when it comes to the manufacturing sector, integration is actually significant between countries X and Y, and thus our initial conclusion would turn out to be inaccurate. Still, overall GVC integration is, as has been summarized in the literature review section of this paper, an established predictor of long-term technology diffusion. Furthermore, its value is enhanced by the fact that it captures cross-sector flows and multiplier effects in a way that other measures do not, a property especially valuable in the case of a general purpose technology that diffuses widely and across all sectors.

Knowledge flows and research collaborations, in turn, do not provide insight into the magnitude of specific transfers or collaborations. For example, a Stack Overflow query used to build a frontier AI model may significantly boost diffusion compared to a query for a narrow AI development. Similarly, co-publication on an advanced LLM model might have a greater impact on diffusion than other AI-related papers. Moreover, while GVC integration captures indirect effects, knowledge flows do not. For instance, if India receives knowledge from the USA and then transfers it to Senegal, it raises the question of whether the transfer originates from the USA, India, or both. This could imply that dependence on the USA and China for research and knowledge transfers is higher than indicated in this study's results.

Another methodological weakness is the difficulty in comparing the three avenues of diffusion, as they are fundamentally different. GVC participation, despite being an established avenue for technology diffusion, is a fairly imprecise predictor due to the age of the data, yet it remains significant for understanding AI diffusion through the economy. In contrast, research and knowledge transfers offer more immediate, micro-level, and AI-specific measurements of diffusion. Although all three are useful for the research goals of this paper, their differences preclude combining them into a common index to measure overall technology diffusion potential, as it would be challenging to assign appropriate weights to each. Consequently, it may be theorized that these measures are not equally disruptive. While this is likely true, it does not necessarily mean that international trade, as a macro measure, will have a greater impact. In fact, in the short term, direct knowledge transfers and research network participation seem to have a larger effect on diffusion than GVC participation. At the same time, long-term efforts to consider these three pathways in tandem should not be abandoned, especially since it seems plausible that interactive effects exist between them.

Another limitation is the use of relatively short and limited timeframes for measuring trends. While data was available for longer periods, additional time resources might have revealed insights that the current analysis does not capture.

Lastly, there is a case for interpreting nominal diffusion rates without adjusting for population, as these represent the actual magnitude of diffusion. However, a high number of requests or collaborations within a smaller population may indicate faster diffusion. For instance, if a country with ten million inhabitants has the same number of fulfilled requests as a country with one hundred million, diffusion in the former is likely faster, as these requests can spread more rapidly among individuals and innovation clusters. Alternatives to population could include the number of researchers for research networks and the number of companies for knowledge transfers, though this data was not readily available.

### 5.3: What are the theoretical and methodological contributions of this paper?

Overall, this paper contributes both theoretically and methodologically in two key ways. Firstly, it provides a comprehensive literature review on the avenues through which technologies can disseminate to LMICs and highlights the importance of this process for economic competitiveness. Global value chains (GVCs) emerge as a long-term pathway, where increased competition and rapid access to new technologies foster innovation and efficiency, leading to knowledge spillovers and productivity gains. Conversely, disembodied diffusion occurs in the short to medium term through various channels such as inter-firm knowledge flows, research, and patents, again resulting in knowledge spillovers that enhance productivity. While LMICs are anticipated to benefit from AI diffusion, trends in AI-driven automation and their exclusion from key international trade agreements pose risks of reduced competitiveness through processes like on-shoring, decreased FDI, and stagnation in innovation. However, deeper GVC integration, enhanced inter-firm knowledge flows, and active participation in research networks have been shown to benefit LMICs, mitigating the aforementioned losses in competitiveness.

Building on the literature review, this paper establishes a framework for better understanding AI diffusion to LMICs through three pathways representing long, medium, and short-term diffusion. GVC integration, research networks, and inter-firm knowledge flows are chosen as proxies for this purpose. Additionally, data availability supports the inclusion of both domestic and international diffusion, enabling a more comprehensive empirical analysis.

**5.4: What are the findings around inter-country diffusion to the chosen countries?**

This paper presents an empirical analysis of current and predicted AI diffusion into sixteen LMICs, examining trends and making comparisons with four developed nations. Regarding GVC integration, a clear and expected tradeoff between backward and forward integration is observed. Countries with high degrees of backward integration (a high percentage of exports containing foreign value added) tend to have weak forward integration (a low percentage of domestic value added in foreign exports). This characteristic will be discussed in more detail below, particularly in the context of whether fast or slow diffusion is beneficial. Another finding is that backward GVC integration, seen as a more likely avenue for international AI diffusion than forward integration, is relatively low with the USA for the countries chosen. Instead, it is generally much higher with China. Based on this, it can be theorized that, in certain areas of international trade, China will play a more dominant role in the diffusion of AI technologies to the investigated countries. A notable finding in this area is the significant degree of integration between Cambodia and Vietnam with China, where tech final and intermediate products play a crucial role. China exports a substantial amount of intermediate products, such as semiconductors, to these two countries for assembly into final products that are then exported. As is commonly known, manufacturing, particularly of computer and similar products, is a sector that both produces a high number of AI innovations and adopts AI technologies to a high degree.

That being said, the imprecise nature of GVC integration, as discussed in section 5.2, inhibits us from drawing definitive conclusions regarding the importance of China versus the USA in being sources of AI diffusion through international trade. A LMIC may hypothetically be deeply backwards integrated with China, while licensing 100% of AI technologies from the US, a process not captured by the measure of GVC integration.

Regarding diffusion through research networks, Tunisia stands out as a major participant and is therefore expected to experience a high degree of AI diffusion through this pathway. Interestingly, this diffusion does not originate from the USA nor China, with each representing only around three percent of co-publications with Tunisia. Overall, the USA plays a more significant role in diffusion through research than through international trade, outperforming China, albeit not by a large margin. Assuming that diffusion through research networks is a more immediate route for accessing AI technologies, and considering the USA's edge in frontier AI models, countries in the sample looking to accelerate diffusion may benefit more from deepening research relations with the USA rather than trade relations with China.

In terms of diffusion through knowledge transfers, Tunisia once again ranks on top, although not as prominently as in research networks. Reliance on the USA for this pathway is very significant, with Nigeria, for instance, having almost 28 percent of its AI-related Stack Overflow requests answered by the USA. China plays an almost non-existent role here, which may be partly due to it lagging behind the USA in this domain but more importantly likely stems from the language barrier and the lesser popularity of the Stack Overflow platform. A further notable insight is that Cameroon ranks next to last in knowledge transfers but seventh in terms of research collaborations, further highlighting the heterogeneity observed not only in the overall rate of diffusion but also across various types of diffusion.

**5.5: How do the sixteen LMICs compare to the developed world, and is the "gap" between them actually closing?**

As anticipated, LMICs significantly lag behind developed nations. In AI research network participation, for instance, Sweden, the top performer, boasts over 500 times the publications per one million inhabitants compared to Myanmar. On average, the four developed countries had 17 times the publications per one million inhabitants compared to the 16 LMICs. A similar disparity is evident in knowledge flows, with the United Kingdom leading with 360 times the knowledge transfers per one million inhabitants compared to Senegal, the lowest-ranking country. On average, developed nations received 11 times the knowledge transfers of the included LMICs.

However, when examining trends, a more optimistic picture emerges. In terms of diffusion through research networks, all LMICs, except Myanmar, experienced substantial increases when comparing the 2018-2020 and 2021-2023 averages, with a growth of 60.1 percent compared to a 25.4 percent increase in developed nations. Although data on AI knowledge flows is less reliable due to limited data points, LMICs also showed greater average increases in this area.

Based on this information, one might argue that the gap between developed and developing nations is closing, and there is some truth to this claim. The data indicates that the rate of diffusion in LMICs is indeed growing faster than in developed countries. Importantly, these figures encompass both international and national AI diffusion. However, it is crucial to consider the context of technology intensities. It is reasonable to assume that technology intensities were initially much higher in developed nations. Given the likely diminishing marginal impact of diffusion on technology intensity, it is unsurprising that increases in diffusion rates are lower for developed nations. Additionally, we lack a comprehensive understanding of how uniformly this relationship applies. Does a one percent increase in diffusion lead to a similar increase in technology intensity across different countries, or are some nations better equipped to absorb and utilize this increase?

While the findings of this paper suggest a potentially narrowing gap, without a more nuanced understanding of the link between diffusion and technology intensities, these conclusions remain tentative.

**5.6: Are all types of diffusion created equal?**

Another crucial question to consider is whether all types of diffusion are created equal. This topic can be approached from multiple perspectives, but this section will focus on just two. First, we explore the differences between the three avenues of diffusion: Which channel drives diffusion the quickest and has the most significant impact on technology intensity? The former question is partially addressed in this paper, identifying international trade as the slowest channel, followed by research networks in the medium term, and knowledge flows in the short term. The latter question, however, remains unanswered as we do not yet know the precise impact a one percent increase in diffusion through global value chains, research networks, and knowledge flows has on technology intensity. Future estimates may reveal that the answer largely depends on the specific country. The unique characteristics of each economy, such as its absorptive capacity of innovation, and other factors will determine which form of diffusion is most impactful. Some countries may have innovation or tax policies that facilitate easy adoption by companies, making direct knowledge transfers and international trade more impactful. Others may already have a robust ecosystem connecting academia and industry, allowing research networks to serve as effective bridges between diffusion and adoption. Moreover, advanced research networks have the advantage of being well-informed about current and upcoming trends, enabling more efficient integration of technologies and the formulation of intentional policies to minimize labor market disruptions.

Another distinction worth discussing is between domestic and international diffusion: Is one more important than the other for AI-aided innovation? A high degree of domestic diffusion can indicate the presence of innovation clusters, which can be advantageous in times of increased de-coupling and protectionism. Conversely, a high degree of international diffusion might be particularly beneficial for the surveyed countries, as it likely involves more advanced AI technologies, such as frontier innovations flowing from leaders like the USA. However, a low level of domestic diffusion coupled with high international diffusion might mean that only a few advanced actors are gaining access to such technologies, leaving out the majority and thereby limiting actual productivity benefits for the country while further exacerbating domestic inequalities. Furthermore, the rapid inflow of disruptive and advanced AI technologies from abroad can pose significant risks to labor markets unprepared for the transition.

The degree of domestic diffusion can lead to differing conclusions depending on the diffusion avenue examined and the stage of diffusion. For instance, among the surveyed countries, Indonesia has the highest and growing percentage of domestic research collaborations at 53.1 percent, followed by Ukraine (47.8 percent) and France (40.0 percent). In contrast, Sweden has a domestic collaboration percentage of only 17.1 percent. For Indonesia, a high percentage may not be advantageous, as it might indicate that the country's researchers are missing out on valuable knowledge transfers through international projects. For Sweden, a small, highly innovative country with a relatively weak AI development landscape, an 82.9 percent share of international collaborations in AI research could reflect an intentional strategy to gain access to knowledge transfers applicable across various sectors of the economy. In the case of France, with domestic actors like Mistral, a high percentage of domestic collaborations likely does not imply missing out on future productivity gains.

Similarly, a high or growing percentage of domestic knowledge flows can indicate a landscape conducive to the widespread adoption of internationally accessed AI technologies, provided the country has access to advanced technologies initially. However, this also poses risks, such as in Indonesia, which has seen a 152.4 percent increase in domestic knowledge transfers when comparing the 2016-2018 and 2019-2021 averages. Indonesia's heavy reliance on manufacturing means that such a substantial increase could signal impending large-scale labor market disruptions due to significant advancements in automation.

**5.7: What rate of diffusion is ideal, and what are the takeaways for AI governance?**

Unsurprisingly, the answer is that it very much depends. Findings of this paper reveal that there is no one-size-fits-all approach to AI diffusion. The current debate often oscillates between those who extol AI's immense productivity benefits and those who advocate for caution, fearing the rapid development and adoption of AI. Research on the trade-offs between these positions is still emerging and is primarily theoretical.

While this debate is crucial, the perspective of LMICs frequently gets overlooked. These countries are not only ill-equipped for rapid AI diffusion but also face the risk of falling further behind developed nations due to their slower access to productivity-enhancing technologies.

Firstly, all long-term global solutions on AI must encompass mechanisms for a more equitable distribution of AI profits. Many LMICs are ill equipped for the swift proliferation of AI technologies compared to developed nations, thereby risking profound economic, political, and societal disruptions. However, slow diffusion also entails analogous risks of economic and societal decline. Thus, global redistribution mechanisms for AI induced economic gains emerge as a fulcrum in the delicate balancing act between slow and rapid diffusion which LMICs will undoubtedly face in the next few decades.

Secondly, international accords aimed at universally decelerating AI development and diffusion are prone to failure not only due to resistance from AI labs and other private companies, but also due to the diverse needs of participating nations. As demonstrated, ideal diffusion rates are not only variable but subject to change, contingent on each nation's continuous assessment of the risks and benefits associated with both rapid and slow AI diffusion. Furthermore, there exists considerable heterogeneity in the specific pathways of AI diffusion that each country might choose to prioritize. Hence an emphasis on tailored and constantly re-examined bi-lateral agreements may serve as a crucial complement to the broader international dialogue on AI governance.

**5.8: What are the next steps for research?**

As highlighted in the above discussion, much remains to be explored in the domain of AI diffusion. Firstly, the link between diffusion rates and technology intensity requires more detailed investigation. Specifically, it is necessary to understand how various types of diffusion impact technology intensities and how these impacts differ across countries.

Furthermore, the relationship between technology intensity and productivity gains warrants further research. Sectoral analyses are also crucial, as it is evident that diffusion occurs more rapidly in certain sectors. For instance, the tech sectors in Cambodia and Vietnam are highly intertwined with China's, likely leading to higher levels of diffusion compared to sectors such as cultural activities.

Concrete next steps would involve conducting case studies of specific LMICs, employing methodologies from this paper and other publications to measure AI diffusion, establish the relationship with technology intensity, and determine ideal diffusion rates based on each country's economic and political characteristics. These studies could help identify optimal diffusion rates to maximize competitiveness while minimizing adverse labor market impacts. Subsequently, this information could be used to assess whether the current and predicted AI governance frameworks in these nations align with ideal diffusion rates and to suggest necessary adjustments for better alignment.

On a more macro level, improved methods for measuring AI diffusion are needed. Surveys should be distributed to firms and public bodies to gauge AI diffusion. These surveys could be integrated with those already used for national accounts, facilitating the creation of input-output satellite accounts that enable a more targeted analysis of international trade as a pathway for AI diffusion.

Lastly, more research needs to be done on expected labor market effects of AI diffusion in LMICs. An interesting debate has emerged regarding whether the non-complementary nature (by AI) of jobs in LMICs or the high percentage of skill intensive jobs being impacted by AI will carry more importance in determining impacts. However, much more work remains to be done in this area, which is sure to carry significant weight in calculations of optimal diffusion rates to LMICs.

## 6. Conclusion

In the wake of rapid AI advancements, this paper aspires to deepen the understanding of AI diffusion, particularly to low-middle-income countries (LMICs), which not only encompass a significant majority of the Earth's population but also possess economies more susceptible to structural changes than those of the developed world.

Firstly, a comprehensive literature review was conducted to enhance the theoretical understanding of AI diffusion and its implications for LMICs. It was discerned that diffusion through global value chains and knowledge flows, represented by international research collaboration, inter-company knowledge transfers, and patenting, constitute the primary conduits for the current and forthcoming dissemination of AI technologies. LMICs appear to be at a disadvantage in the new AI era, facing risks such as loss of competitiveness due to increased AI-aided automation in manufacturing. However, greater engagement of LMICs in GVCs, research networks, and corporate clusters has been shown to augment technology diffusion and productivity, potentially mitigating adverse economic impacts.

Secondly, an empirical analysis was conducted across three dimensions: GVC integration, AI research network participation, and direct AI-related knowledge transfers, encompassing 16 LMICs and making comparisons with four advanced economies. Additionally, the dependence on the USA and China for diffusion was scrutinized. While some conclusions were drawn regarding individual countries, it became apparent that current AI diffusion in the 16 countries is markedly lower than in the developed world, though a trend towards a narrowing gap was identified. The significance of China for diffusion through global value chains contrasts with the greater role of the USA in diffusion through research and knowledge transfers.

The paper has certain limitations, including the omission of potentially valuable data on patents and FDI, additional avenues for AI diffusion, and the exclusion of low-income and more developed nations due to data and time constraints. The methodologies employed are somewhat constrained by their incompatibility with each other in representing different timeframes and scales of diffusion. They also fail to account for indirect paths for research and knowledge transfers and do not include LLMs as a potential source for diffusion. Furthermore, specifically GVC integration lacks a sectoral granularity which would have allowed for a more precise understanding of this avenue. Finally trends, while examined, are temporally limited.

The study provides additional value by raising macro-level questions, such as whether the gap between LMICs and developed nations is closing, whether all types of diffusion are equally beneficial, and what rates of diffusion are optimal. While trends suggest positive developments for LMICs, implying a narrowing gap, the methods used cannot conclusively measure the relationship between diffusion and technology intensity, and thus productivity gains. Therefore, definitive conclusions cannot be drawn. The effectiveness of various diffusion avenues depends entirely on the specific country's economic strengths and weaknesses. The same applies to the distinction between domestic and international diffusion, where both high and low degrees can be advantageous or detrimental depending on the economy's characteristics. Lastly, regarding the optimal rate of diffusion, LMICs face a conundrum: slow diffusion rates may safeguard against large-scale labor market disruptions but lead to a loss of competitiveness, whereas high rates may enhance competitiveness but risk labor market destabilization.

Much remains to be researched in AI diffusion, particularly the link between diffusion rates and technology intensities, and a more nuanced understanding of sectoral idiosyncrasies. Case studies of LMICs are needed to calculate optimal diffusion rates based on a meticulous risk-benefit analysis, which can then be used to evaluate their alignment with current and predicted AI governance frameworks. On a macro level, more precise methodologies for measuring AI diffusion need to be established, and more in-depth research is needed on the particularities of labor markets in LMICs, in contrast to developed economies.

Meanwhile, international governance of AI must consider the balancing and risk reducing nature of global redistribution mechanisms for AI-induced economic gains, and the value of bilateral agreements for tackling the diverse needs and corresponding risks faced by economies transitioning into an AI-dominated era.